\documentclass[12pt]{article}

\usepackage{graphicx}
\usepackage{amssymb}
\begin{document}

\newcommand{\vecx}{\mbox{\boldmath $x$}}


%
\begin{center}
{\large\bf
Nonextensive quantum method for
itinerant-electron ferromagnetism:
Fractorization approach 
} 
\end{center}

\begin{center}
Hideo Hasegawa
\footnote{hideohasegawa@goo.jp}
\end{center}

\begin{center}
{\it Department of Physics, Tokyo Gakugei University,  \\
Koganei, Tokyo 184-8501, Japan}
\end{center}
\begin{center}
({\today})
\end{center}
\thispagestyle{myheadings}

\begin{abstract}
Magnetic and thermodynamical properties of itinerant-electron
(metallic) ferromagnets described by the Hubbard model
have been discussed with the use of the generalized Fermi-Dirac (GFD) 
distribution for nonextensive quantum systems.
We have derived the GFD distribution within the superstatistics, 
which is equivalent to that obtained by the maximum-entropy 
method to the Tsallis entropy with the factorization approimation.
By using the Hartree-Fock approximation
to the electron-electron interaction in the Hubbard model, 
we have calculated magnetic moment, energy, specific heat 
and Curie-Weiss-type spin susceptibility, as functions of 
the temperature and entropic index $q$ expressing the degree of
the nonextensivity: $q=1.0$ corresponds to the Boltzmann-Gibbs statistics.
It has been shown that with increasing the nonextensivity of $\vert q-1 \vert$,
the temperature dependence of magnetic moment becomes more
significant and the low-temperature electronic specific heat is much increased.
This is attributed to enlarged Stoner excitations in
the GFD distribution, which is elucidated by
an analysis with the use of the generalized Sommerfeld expansion.
We discuss the difference and similarity between the effects of
the nonextensivity on metallic and insulating ferromagnets.

\end{abstract}

\noindent
\vspace{0.5cm}

\vspace{1cm}

\noindent

 

\newpage

\section{Introduction}

Since Tsallis proposed the nonextensive statistics in 1988
\cite{Tsallis88}, considerable works on the related study
have been made (for a recent review, see \cite{Tsallis04}).
It is based on the generalized entropy (called the Tsallis
entropy) which is a one-parameter generalization of the Boltzmann-Gibbs 
entropy with the entropic index $q$: the Tsallis entropy
in the limit of $q=1.0$ reduces to the Boltzmann-Gibbs entropy.
The nonextensive statistics has been successfully applied 
to a wide class of subjects including physics, chemistry, information science,
biology and economics \cite{NES}.
Despite elegant formalism of the Tsallis nonextensive statistics,
there are four possible methods in an evaluation of expectation values 
with the maximum-entropy method (MEM):
(i) original method \cite{Tsallis88},
(ii) unnormalized method \cite{Curado91},
(iii) normalized method \cite{Tsallis98} and
(iv) the optimal Lagrange multiplier method \cite{Martinez00},
although the four methods are equivalent in the sense that distributions 
derived in them are easily transformed each other \cite{Ferri05}.
A comparison among the four methods is made in Ref. \cite{Tsallis04}.

An alternative approach to nonextensive systems is the superstatistics 
\cite{Wilk00,Beck01}. Complex nonextensive systems are expected to 
temporary and spatially fluctuate. It is assumed that locally 
the equilibrium state is described by the Boltzmann-Gibbs statistics, 
and that their global properties may be expressed 
by a superposition of them over some intensive physical quantity,
{\it e.g.} the inverse temperature \cite{Wilk00}-\cite{Beck07}.
Many applications of the concept of the superstatistics have been 
pointed out (for a recent review, see \cite{Beck07}).
It is, however, not clear how to obtain the mixing probability 
distribution of fluctuating parameter from first principles.
This problem is currently controversial and
some attempts in this direction have been proposed
\cite{Souza03}-\cite{Straeten08}.

The nonextensive statistics has been applied not only to
classical systems but also to quantum ones 
\cite{Tsallis95}-\cite{Hasegawa07}. 
For fermion systems,
the generalized Fermi-Dirac (GFD) distribution 
was derived by the asymptotic approach for 
$ \vert q-1 \vert/k_B T \rightarrow 0$ \cite{Tsallis95}
and by the MEM (ii) with the factorization approximation \cite{Buy95}.
With the use of the MEM (ii) 
and MEM (iii), 
Refs. \cite{Rajagopal98,Lenzi99} have derived
the formally exact expression for the grand canonical partition
function of nonextensive systems, which
is expressed as a contour integral in terms of
the Boltzmann-Gibbs counterpart.
Although the exact formulation is very valuable, the actual calculation of
the contour integral is difficult and it may be performed only 
in the limited cases at the moment \cite{Rego03}.
Quite recently, the nonextensive quantum extension
has been proposed by using the MEM (iv) \cite{Cavallo08}. 
Among the three approaches of the asymptotic \cite{Tsallis95}, 
factorization \cite{Buy95} and exact methods \cite{Rajagopal98,Lenzi99}
in nonextensive quantum statistics, the factorization approach is 
the easiest way for handling physical systems \cite{Wang97,Torres98}. 
The nonextensive quantum statistics has been applied to various
subjects including black-body radiation \cite{Tsallis95,Wang98},
Bose-Einstein condensation \cite{Torres98,Torres00,Biswas06,Biswas08},
metallic \cite{Oliveira00} and superconducting 
materials \cite{Nunes01,Uys01}, 
spin systems \cite{Portesi95,Nobre96,Reis02,Reis03,Reis06}
and nano-magnetism \cite{Hasegawa05,Hasegawa07}.

Now we pay our attention to magnetic systems.
Although there are many magnetic materials,
they are classified into two categories:
insulating and metallic magnets.
The latter are often referred to also
as itinerant-electron, collective-electron or band magnets
({\it metallic} and {\it itinerant-electron} are interchangeably used 
hereafter).
In insulating magnets, for example, 
of rare-earth elements such as Gd and La,
f electrons form the localized spin at each atomic site
which yields integer magnetic moment in units of $\mu_B$
(Bohr magneton). 
On the contrary, in itinerant-electron magnets of transition metals
such as Fe, Co and Ni, d electrons not only form magnetic
moment at each atomic site 
but also itinerate in crystals, by which 
materials become metallic.
The magnetic moment of itinerant-electron magnets
is not integer in units of $\mu_B$.
A modern theory of magnetism has a long history 
over the last half century since the advent of quantum mechanics.
Insulating magnets are well described by the Heisenberg
model. With the Weiss molecular-field theory
and more advanced theories for the Heisenberg model,
our understanding of magnetic
properties such as magnetic structures and phase transition
has been much deepen. On the other hand,  
a study of itinerant-electron magnets
was initiated by Stoner \cite{Stoner36} 
and Slater \cite{Slater36}.
Later Hubbard proposed the so-called 
Hubbard model \cite{Hubbard64}, which has been widely
adopted for a study of itinerant-electron magnetism.
Studies with the Hartree-Fock (mean-field) approximation 
to the electron-electron interaction in the Hubbard model
account for the non-integral magnetic moment
and the large $T$-linear coefficient of the specific heat
at low temperatures, which are experimentally observed.

Nonextensive statistics has been applied to insulating ferromagnets,
by using the MEM (ii) \cite{Portesi95}\cite{Nobre96}
and the MEM (iii) \cite{Reis02}-\cite{Hasegawa05}.
Peculiar magnetic properties observed in manganites are reported 
to be well accounted for by the nonextensive statistics
\cite{Portesi95}-\cite{Reis06}.
The purpose of the present paper is to apply the nonextensive quantum statistics
to  itinerant-electron ferromagnets with the use of the GFD distribution
derived within the superstatistics.
The resultant GFD distribution is equivalent
to that obtained by the MEM (ii) with the factorization approximation 
\cite{Buy95}, which is valid for dilute fermion gas
\cite{Pennini95,Wang97,Torres98}. 
The factorized GFD distribution \cite{Buy95}
has been applied to various quantum subjects
\cite{Torres00,Biswas06,Biswas08,Oliveira00,Nunes01,Uys01}
because it is a good, practical approximation \cite{Wang97,Torres98}.
We have calculated the magnetic moment, energy, specific heat and 
Curie-Weiss-type susceptibility of itinerant-electron ferromagnets 
described by the Hubbard model with the Hartree-Fock approximation.
Such a calculation is worthwhile, clarifying the effect of the nonextensivity
on metallic ferromagnets which is different from that on insulating counterparts. 
Our study is the first application of the nonextensive quantum
method to itinerant-electron ferromagnets, as far as we are aware of.

The paper is organized as follows. In Section 2, we discuss 
the adopted Hubbard model and the GFD distribution
derived by the superstatistics.
Analytical expressions for magnetic moments, 
energy, specific heat and
susceptibility are presented with some model calculations.
In Section 3, qualitative discussions 
on magnetic and thermodynamical properties are made
with the use of the generalized Sommerfeld low-temperature 
expansion for physical quantities.
Relevance of our calculation to heterogeneous 
magnets such as metallic spin glass and amorphous metals 
is discussed.
Section 4 is devoted to our conclusion. 

\section{Formulation}
\subsection{Adopted model}

We have considered itinerant-electron
ferromagnets described by the Hubbard model given by
\cite{Hubbard64}
\begin{eqnarray}
\hat{H} &=& \sum_{\sigma}\sum_i \epsilon_0 \:n_{i \sigma} 
+\sum_{\sigma} \sum_{i,j} t_{ij} \:a^{\dagger}_{i \sigma}a_{j \sigma}
+ U \:\sum_i n_{i\uparrow}n_{i\downarrow}
-\mu_B B \:\sum_i(n_{i\uparrow}-n_{i\downarrow}),
\label{eq:A1}
\end{eqnarray}
where $n_{i \sigma}=a^{\dagger}_{i \sigma}a_{i \sigma}$,
$a_{i \sigma}$ ($a^{\dagger}_{i \sigma}$) denotes  
an annihilation (creation) operator of 
a $\sigma$-spin electron ($\sigma=\uparrow,\:\downarrow $)
at the lattice site $i$, $\epsilon_0$ the intrinsic energy of atom,
$t_{ij}$ the electron hopping, $U$ the intra-atomic electron-electron
interaction and $B$ an applied magnetic field.
We have adopted the Hartree-Fock approximation to the electron-electron
interaction of the third term in Eq. (\ref{eq:A1}), as given by
\begin{eqnarray}
U n_{i \uparrow} n_{i \downarrow} 
&\simeq&  U \langle n_{\uparrow} \rangle \: n_{i \downarrow}
+ U \langle n_{ \downarrow} \rangle \: n_{i\uparrow }
- U \langle n_{ \uparrow} \rangle\: \langle n_{ \downarrow}\rangle,
\label{eq:A2}
\end{eqnarray}
where $ \langle n_{\sigma} \rangle $
denotes the average of number of electrons with
spin $\sigma$ to be evaluated shortly 
[see Eqs. (\ref{eq:C1}) and (\ref{eq:C2})].
With the Hartree-Fock approximation, Eq. (\ref{eq:A1}) becomes 
the effective one-electron Hamiltonian given by
\begin{eqnarray}
\hat{H} &\simeq& \hat{H}_{\uparrow} + \hat{H}_{\downarrow}
- U \langle n_{ \uparrow} \rangle\: \langle n_{ \downarrow}\rangle,
\label{eq:A3}
\end{eqnarray}
where 
\begin{eqnarray}
\hat{H}_{\sigma} &=& \sum_i \epsilon_{\sigma} n_{i \sigma} 
+ \sum_{i,j} t_{ij} a^{\dagger}_{i \sigma}a_{j \sigma}
\hspace{1cm}\mbox{($\sigma= \uparrow$ and $\downarrow$)}, 
\label{eq:A4}
\end{eqnarray}
with
\begin{eqnarray}
\epsilon_{\uparrow,\downarrow} &=& \epsilon_0 
+ U n_{\downarrow} \mp \mu_B B,
\label{eq:A5} 
\end{eqnarray}
the minus (plus) sign in Eq. (\ref{eq:A5}) being applied 
to $\uparrow$-spin ($\downarrow$-spin) electrons.

\subsection{GFD distribution within the superstatistics}

We have considered the nonextensive fermion system consisting of
many clusters, each of which includes $N$ particles with total energy $E$. 
Non-equilibrium or quasi-equilibrium states of the system
are expected to temporary and spatially fluctuate.
However, locally the equilibrium state of a given cluster is assumed to
be described by the Boltzmann-Gibbs statistics in the
superstatistics \cite{Wilk00}-\cite{Beck05}.
The probability distribution for $\{ n_k \}$ 
in a given cluster with the local temperature $\tilde{T}$ is given by
\begin{eqnarray}
p_{BG}(\tilde{\beta},\{ n_k \}) 
&=& \frac{1}{\Xi(\tilde{\beta})} 
\prod_k e^{-\tilde{\beta}(\epsilon_k-\mu)n_k},
\label{eq:H1}
\end{eqnarray}
where the grand-partition function $\Xi(\tilde{\beta})$ is given by
\begin{eqnarray}
\Xi(\tilde{\beta}) &=& \prod_k [1+e^{-\tilde{\beta}(\epsilon_k-\mu)}].
\label{eq:H2}
\end{eqnarray}
Here $\tilde{\beta}=1/k_B \tilde{T}$, $k_B$ is the Boltzmann constant, and
$n_k$ and $\epsilon_k$ denote the number of states
and energy, respectively, of the state $k$. 

After the concept of the superstatistics \cite{Wilk00}-\cite{Beck05},
we have assumed that the inverse of the temperature ($\tilde{\beta}$)
fluctuates and its distribution is given by 
the $\chi^2$-distribution with rank $n$ given by
\begin{eqnarray}
g(\tilde{\beta}) &=& \frac{1}{\Gamma(\frac{n}{2})} 
\left( \frac{n}{2 \beta} \right)^{\frac{n}{2} } 
\tilde{\beta}^{\frac{n}{2}-1} e^{- \frac{n\tilde{\beta}}{2 \beta} },
\label{eq:H4}
\end{eqnarray}
where $\Gamma(x)$ denotes the gamma function.
The average and variance of $\tilde{\beta}$ are given by
$ \langle \tilde{\beta} \rangle_{\tilde{\beta}}=\beta$
and $(\langle \tilde{\beta}^2 \rangle_{\tilde{\beta}}-\beta^2)/\beta^2=2/n$, 
respectively. 
The distribution averaged over the system 
with the temperature $T$ ($=1/k_B \beta$) 
is assumed to be given by
\begin{eqnarray}
p(\beta, \{ n_k\}) &=& 
\int_0^{\infty} p_{BG}(\tilde{\beta},\{ n_k \}) 
\:g(\tilde{\beta})\:d \tilde{\beta}.
\label{eq:H3}
\end{eqnarray}
When we adopt the type-A superstatistics 
in which the $\tilde{\beta}$ dependence of $\Xi(\tilde{\beta})$ 
is neglected \cite{Beck07}, 
Eqs. (\ref{eq:H4}) and (\ref{eq:H3}) yield
\begin{eqnarray}
p(\beta,\{ n_k\}) &=& \frac{1}{Z_q}
\prod_k \exp_q[-\beta(\epsilon_k-\mu)n_k],
\label{eq:H5}
\end{eqnarray}
with
\begin{eqnarray}
Z_q &=& \prod_k \{1+\exp_q[-\beta(\epsilon_k-\mu)]\}, 
\label{eq:H6}\\
q &=& \frac{2}{n}+1,
\label{eq:H7}
\end{eqnarray}
where $\exp_q(x)$ stands for the $q$-exponential function
defined by
\begin{eqnarray}
\exp_q(x) &=& [1+(1-q)x]^{\frac{1}{1-q} }
\hspace{1cm}\mbox{for $1+(1-q)x \geq 0$}, 
\label{eq:B8}\\
&=& 0
\hspace{4cm}\mbox{for $1+(1-q)x < 0$}.
\label{eq:B9}
\end{eqnarray}
Equation (\ref{eq:B9}) expresses the cut-off properties of the
$q$-exponential function.

The probability of occupation of the state $k$ by $n_k$ particles
is given by 
\begin{eqnarray}
P(n_k) &=&
\frac{[1-(1-q)\beta(\epsilon-\mu)n_k]^{\frac{1}{1-q} }}{Z_q} \nonumber \\
&& \times 
\prod_{j\neq k} \{1+[1-(1-q) \beta(\epsilon_j-\mu)]^{\frac{1}{1-q} } \}\\
&=&\frac{[1-(1-q)\beta(\epsilon-\mu)n_k]^{\frac{1}{1-q}}}
{1+[1-(1-q) \beta(\epsilon_k-\mu)]^{\frac{1}{1-q}}},
\end{eqnarray}
where the factorization approximation is employed \cite{Buy95}.
The probability of occupation of the quantum state with 
the energy $\epsilon_k$, $f_q(\epsilon_k)$, is given by
\begin{eqnarray}
f_q(\epsilon_k) &=& 
\sum_{n_k} P(n_k) n_k,\\
&=& \frac{1}{1+\left( \exp_q[-\beta(\epsilon_k-\mu)] \right)^{-1}}.
\label{eq:H10} 
\end{eqnarray}
Averages of the number of electrons and energy per cluster
are expressed in terms of $f_q(\epsilon_k)$ by
\begin{eqnarray}
N &=& \sum_k \: f_q(\epsilon_k),
\label{eq:B10} \\
E &=& \sum_k \: f_q(\epsilon_k) \:\epsilon_k.
\end{eqnarray}

The GFD distribution given in Eq. (\ref{eq:H10}) is equivalent 
to that obtained by the MEM (ii) with the factorization approximation 
for $0 < q < 2$ \cite{Buy95}.
In the present paper,
we will adopt the GFD distribution given by Eq. (\ref{eq:H10})
not only for $q \geq 1.0$ but also $q \leq 1.0$,
although it is valid for $q \geq 1.0$
within the superstatistics [Eq. (\ref{eq:H7})].
One of the advantages of the superstatistics is
that the entropic index $q$ is expressed in terms of model
parameters as given by Eq. (\ref{eq:H7})
(related discussion being given in Section 3).

In the limit of $n \rightarrow \infty$ where
$g(\tilde{\beta})=\delta(\tilde{\beta}-\beta)$ [Eq. (\ref{eq:H4})] and
$q=1.0$ [Eq. (\ref{eq:H7})], Eq. (\ref{eq:H10}) reduces
to the conventional Fermi-Dirac distribution obtained
in the Boltzmann-Gibbs statistics,
\begin{eqnarray}
f_1(\epsilon_k) &=& f_{BG}(\epsilon_k)
= \frac{1}{1+e^{\beta(\epsilon_k-\mu)}}.
\label{eq:H11}
\end{eqnarray}

In the zero-temperature limit of $\beta \rightarrow \infty$, 
both Eqs. (\ref{eq:H10}) and (\ref{eq:H11}) reduce to
\begin{eqnarray}
f_q(\epsilon_k) &=& f_{1}(\epsilon_k)
= \Theta(\mu-\epsilon_k)
\hspace{2cm}\mbox{for $\beta \rightarrow \infty$},
\label{eq:B11}
\end{eqnarray}
where $\Theta(x)$ denotes the Heaviside function:
$\Theta(x)=1$ for $x > 0$ and zero otherwise. 
Equation (\ref{eq:B11}) shows that the quantum state at $T=0$
is not modified by the nonextensivity \cite{Torres98}.
The derivative of $f_q(\epsilon)$ with respect to $\epsilon$
is given by
\begin{eqnarray}
\frac{\partial f_q(\epsilon)}{\partial \epsilon}
&=& - \frac{\beta \left(\exp_q[-\beta (\epsilon-\mu)] \right)^{q-2} }
{ \{1+\left(\exp_q[-\beta (\epsilon-\mu)] \right)^{-1} \}^2 },
\label{eq:B13} \\
&\rightarrow & - \frac{\beta e^{\beta (\epsilon-\mu)}}
{[1+e^{\beta (\epsilon-\mu)}]^2}
\hspace{1cm}\mbox{for $q \rightarrow 1.0$}.
\label{eq:B14}
\end{eqnarray}

In the high-temperature limit of $\beta \rightarrow 0$,
Eq. (\ref{eq:H10}) becomes
\begin{eqnarray}
f_q(\epsilon_k) &\simeq& \exp_q(-\beta \epsilon_k)
\hspace{2cm}\mbox{for $\beta \rightarrow 0$},
\end{eqnarray}
with $\mu=0$.

Figs. \ref{fig1}(a) and (b) show the $\epsilon$ dependence
of $f_q(\epsilon)$ and 
$- \partial f_q(\epsilon)/\partial \epsilon$, respectively,
for various $q$ values with $\beta=1.0$.
When $q$ is increased (decreased) from $q=1.0$, the distribution at
$\epsilon < \mu $ and $\epsilon > \mu $
is increased (decreased). 
The effect of $q$ on the GFD distribution
is more clearly seen in its derivative of 
$\partial f_q(\epsilon)/\partial \epsilon$, which has a power-law
tail at $\vert \epsilon- \mu \vert \gg 1$ for $q \neq 1.0$.
Because of the cut-off properties of the $q$-exponential function 
given by Eq. (\ref{eq:B9}), we obtain
\begin{eqnarray}
f_q(\epsilon) &=& 0.0, \;\;
\frac{\partial f_q(\epsilon)}{\partial \epsilon}=0
\hspace{0.5cm}\mbox{at $(\epsilon-\mu) 
>  \frac{1}{(1-q)\beta}$ for $q < 1$},
\nonumber \\
f_q(\epsilon) &=& 1.0, \;\;
\frac{\partial f_q(\epsilon)}{\partial \epsilon}=0
\hspace{0.5cm}\mbox{at $(\epsilon-\mu) 
< -\frac{1}{(q-1)\beta}$ for $q > 1$}, 
\label{eq:B16}
\end{eqnarray}
which is clearly realized in Fig. \ref{fig1}.

\subsection{Magnetic moment}

By using Eqs. (\ref{eq:A3}), (\ref{eq:H10}) and (\ref{eq:B10})
with $\epsilon_0+Un/2=0$, 
we obtain the self-consistent equations
for the magnetic moment ($m$) and the number of
electrons ($n$) per lattice site, given by
\begin{eqnarray}
m &=& n_{\uparrow}-n_{\downarrow}
= \int [\rho_{\uparrow}(\epsilon)-\rho_{\downarrow}(\epsilon)]
f_{q}(\epsilon)\:d\epsilon, 
\label{eq:C1} \\
n &=& n_{\uparrow}+n_{\downarrow}
=\int [\rho_{\uparrow}(\epsilon)+\rho_{\downarrow}(\epsilon)]
f_{q}(\epsilon)\:d\epsilon, 
\label{eq:C2}
\end{eqnarray}
with
\begin{eqnarray}
\rho_{\uparrow, \downarrow}(\epsilon) 
&=& \rho_0\left( \epsilon \pm \left[\frac{Um}{2} 
+ \mu_B B\right] \right),
\label{eq:C3} \\
\rho_0(\epsilon) &=& \frac{1}{N_a} \sum_k \delta(\epsilon-\epsilon_k),
\label{eq:C4}
\end{eqnarray}
where $\rho_0(\epsilon)$ denotes the density of states and
$N_a$ the number of lattice sites:
the plus and minus signs in Eq. (\ref{eq:C3}) are applied to
$ \uparrow$- and $ \downarrow$-spin electrons, respectively. 
From Eqs. (\ref{eq:C1})-(\ref{eq:C4}), $m$ and $\mu$ are 
self-consistently determined
as a function of $T$ for given parameters of
$q$, $n$ and $U$ and density of state, $\rho_0(\epsilon)$.

Bearing Fe, a typical transition-metal ferromagnet, 
in mind, we have performed model calculations
with a bell-shape density of states 
for a single band given by
\begin{eqnarray}
\rho_0(\epsilon) &=& \left( \frac{2}{\pi W} \right)
\sqrt{1-\left(\frac{\epsilon}{W}\right)^2} 
\;\Theta(W- \vert \epsilon \vert),
\label{eq:C5}
\end{eqnarray}
where $W$ denotes a half of the total bandwidth.
It has been reported that $U \sim 2W \simeq 5 $ eV for Fe 
\cite{Hasegawa79,Callaway97}.
Fig. \ref{fig2} shows the ground-state magnetic moment
as a function of $U$ for $n=1.4$ electrons. 
We have employed $U/W=1.75$ which leads to magnetic moment 
of $m=0.47\;\mu_B$ at $k_B T/W=0.0$. 
Adopted values of $n=1.4$ electrons and
$m=0.47 \:\mu_B$ roughly correspond to 
those of Fe which has seven d electrons and 
the ground-state magnetic moment of 2.2 $\mu_B$
({\it i.e.,} 7.0/5=1.40 electron and 2.2/5=0.44 $\mu_B$ per orbital). 
It is noted that the $U$-$m$ relation shown in Fig. \ref{fig2}
is valid for $0 < q < 2$ because $f_q(\epsilon)$ is
independent of $q$ at $k_B T/W=0$ [Eq. (\ref{eq:B11})].
We have solved self-consistent equations (\ref{eq:C1})-(\ref{eq:C4}) 
by changing  $q$ and $T$ with the use of 
the Newton-Raphson method, which is indispensable 
in our calculations, in particular for $q \leq 0.4$ and $q \geq 1.6$
(see the appendix).

Figs. \ref{fig3} (a) and (b) show the temperature 
dependence of the magnetic moment, $m$, for $q \leq 1.0$
and $q \geq 1.0$, respectively.
With increasing $\vert q-1 \vert$, the temperature dependence of
magnetic moments becomes more significant and the
Curie temperature becomes lower.
This is more clearly seen in Fig. \ref{fig4}, where
$T_C$ is plotted as a function of $q$.
The $q$-$T_C$ plot is almost symmetric with respect to $q=1.0$
where we obtain the maximum value of $k_B T_C/W=0.143$.
If we adopt $W \simeq 2.5$ eV obtained by the band-structure calculation 
for Fe \cite{Callaway97}, the calculated Curie temperature 
at $q=1.0$ is $T_C \simeq 3500$ K, 
while the observed $T_C$ of Fe is 
$1044$ K \cite{Crangle71}.

\subsection{Energy and Specific heat}

We calculate the energy per lattice site given by
\begin{eqnarray}
E &=& \int \epsilon 
\: [\rho_{\uparrow}(\epsilon)+\rho_{\downarrow}(\epsilon)]
f_q(\epsilon)\: d\epsilon 
- \frac{U}{4}(n^2-m^2),
\label{eq:D1}
\end{eqnarray}
from which the specific heat is given by
\begin{eqnarray}
C &=& \frac{d E}{dT}=\frac{\partial E}{\partial T}
+ \frac{\partial E}{\partial m} \frac{d m}{d T}
+ \frac{\partial E}{\partial \mu} \frac{d \mu}{d T},
\label{eq:D2}
\end{eqnarray}
with
\begin{eqnarray}
\frac{\partial E}{\partial T}
&=& - \frac{1}{T} \int \epsilon\:(\epsilon-\mu)
[\rho_{\uparrow }(\epsilon)+\rho_{\downarrow}(\epsilon)] 
\:\frac{\partial f_q(\epsilon)}{\partial \epsilon} \: d\epsilon, 
\label{eq:D3} \\
\frac{\partial E}{\partial m}
&=&- \frac{U}{2} \int \epsilon
\:[\rho_{\uparrow }(\epsilon)-\rho_{\downarrow}(\epsilon)] 
\:\frac{\partial f_q(\epsilon)}{\partial \epsilon} 
\: d\epsilon, 
\label{eq:D4} \\
\frac{\partial E}{\partial \mu}
&=& -\int \epsilon
\: [\rho_{\uparrow }(\epsilon)+\rho_{\downarrow}(\epsilon)] 
\:\frac{\partial f_q(\epsilon)}{\partial \epsilon} 
\: d\epsilon.
\label{eq:D5}
\end{eqnarray}
Analytic expressions for $d m/dT$ and $d\mu/dT$ in Eq. (\ref{eq:D2}) 
are expressed by Eqs. (\ref{eq:X9})-(\ref{eq:X10}) 
(for details see the appendix).

Figs. \ref{fig5} (a) and (b) show the temperature 
dependence of the specific heat $C$ for $q \leq 1.0$ and $q \geq 1.0$,
respectively.
With increasing $\vert q-1.0 \vert$, 
the specific heat at low temperatures is increased and 
its temperature dependence is considerably modified. 

\subsection{Spin susceptibility}

The spin susceptibility is expressed by
\begin{eqnarray}
\chi &=& \frac{d m}{d B}, 
\label{eq:E1}
\end{eqnarray}
from which the paramagnetic spin susceptibility is given by
\begin{eqnarray}
\chi &=& \mu_B^2 \:\frac{2 \chi_{0}}{(1-U \chi_{0})},
\label{eq:E2}
\end{eqnarray}
with
\begin{eqnarray}
\chi_0 &=& -\int  \: \rho(\epsilon) 
\frac{\partial f_q(\epsilon)}{\partial \epsilon}\:d\epsilon.
\label{eq:E3}
\end{eqnarray}

The temperature dependence of the inverse of calculated
susceptibility, $1/\chi$,
for $q \leq 1.0$ and $q \geq 1.0$ is shown 
in Figs. \ref{fig6}(a) and (b), respectively.
The Curie temperature $T_C$, which is realized at $1/\chi=0$,
is decreased with increasing $\vert q-1 \vert$, as shown 
in Fig. \ref{fig4}.

Fig. \ref{fig7} shows the Curie temperature $T_C$
as a function of $U$ for various $q$, which are determined
by the divergence of the susceptibility.
The Curie temperature vanishes at $U/W \leq 1.66$ independently
of $q$ because the GFD does not depend on $q$ [Eq. (\ref{eq:B11})].
The Curie temperature is lowered with increasing $\vert q-1 \vert$
(Fig. \ref{fig4}).

\section{Discussion}

It is possible to qualitatively elucidate the
magnetic and thermodynamical properties of nonextensive 
itinerant-electron ferromagnets presented in the preceding section,
with the use of the generalized Sommerfeld expression
of various quantities. 
The integral $I$ including an arbitrary function $\phi(\epsilon)$
and the GFD distribution $f_q(\epsilon)$ 
is given by
\begin{eqnarray}
I &=& \int  \: \phi(\epsilon) f_q(\epsilon)\: d\epsilon, \\
&=& \int^{\mu}  \: \phi(\epsilon)\:d\epsilon
+ \sum_{n=1}^{\infty} \:c_n \:T^n \phi^{(n-1)}(\mu),
\label{eq:F1}
\end{eqnarray}
with
\begin{eqnarray}
c_n &=& - \frac{1}{n!} \int_{-\infty}^{\infty}  
\:x^n\:\frac{d}{dx} \left( \frac{1}{1+[\exp_q(-x)]^{-1}} \right)\:dx,
\label{eq:F2}
\end{eqnarray} 
which is valid at low temperatures. 
Expansion coefficients for $q=1.0$
are given by $c_2=\pi^2/6$ (=1.645), $c_4=7 \pi^4/360$ (=1.894), 
and $c_n=0.0$ for odd $n$.

The $q$ dependence of $c_n$ for $n=1$ to 4
is plotted in Fig. \ref{fig8}, which shows the followings: 
(1) $c_1$ and $c_3$ are not zero for $q \neq 1.0$ \cite{Oliveira00}
(though a magnitude of $c_1$ is small)
in contrast with $c_1=c_3=0$ in the conventional Fermi-Dirac distribution,
(2) with increasing $\vert q-1 \vert $, $c_2$, $\vert c_3 \vert$ 
and $c_4$ are much increased, and
(3) the $q$ dependence of $c_2$ and $c_4$ are almost symmetric 
with respect to $q=1.0$ 
while those of $c_1$ and $c_3$ are nearly anti-symmetric.
The obtained $q$ dependence of $c_n$ may be understood as follows.

When we expand $f_q(\epsilon)$ given by Eq. (\ref{eq:H10}) 
in a series of $(q-1)$ \cite{Torres98}, we obtain
\begin{eqnarray}
f_q(\epsilon) &=& f_{1}(\epsilon) 
+\frac{(q-1)}{2}\beta^2 (\epsilon-\mu)^2 e^{\beta(\epsilon-\mu)} 
f_1(\epsilon)^2+\cdot\cdot,
\label{eq:F3} \\
&=& f_1(\epsilon)-\frac{(q-1)}{2}\beta(\epsilon-\mu)^2 
\frac{\partial f_1(\epsilon)}{\partial \epsilon}+ \cdot\cdot.
\label{eq:F4}
\end{eqnarray}
Substituting Eq. (\ref{eq:F4}) to Eq. (\ref{eq:F1}) and
using the integral by part, we obtain $I$ 
given by
\begin{eqnarray}
I &=& \int \phi(\epsilon) f_1(\epsilon) \:d \epsilon 
+ \frac{(q-1)}{2 T} 
\int [2(\epsilon-\mu)\phi(\epsilon)+(\epsilon-\mu)^2 \phi'(\epsilon)] 
f_1(\epsilon) \:d \epsilon.
\label{eq:F5}
\end{eqnarray}
By using Eq. (\ref{eq:F1}) for the second term of Eq. (\ref{eq:F5}), 
we obtain
\begin{eqnarray}
c_1(q) &=& \frac{(q-1)\pi^2}{6}+\cdot,\\
c_2(q) &=& c_2(1) + O((q-1)^2), \\
c_3(q) &=& \frac{7(q-1)\pi^4}{60}+\cdot,\\
c_4(q) &=& c_4(1) + O((q-1)^2),
\end{eqnarray}
where contributions of $O(q-1)$ to $c_2$ and $c_4$ are vanishing.
Thus the $q$ dependence of $c_2$ and $c_4$ is almost
symmetric with respect to $q=1.0$ whereas that of
$c_1$ and $c_3$ is nearly anti-symmetric, as Fig. \ref{fig9} shows.  

Setting $\phi(\epsilon)=\rho_{\uparrow}(\epsilon)
-\rho_{\downarrow}(\epsilon)$,
$\phi(\epsilon)=\rho_{\uparrow(\epsilon)}
+\rho_{\downarrow}(\epsilon)$, and
$\phi(\epsilon)=\epsilon \: [\rho_{\uparrow(\epsilon)}
+\rho_{\downarrow}(\epsilon)]$ 
in Eq. (\ref{eq:F1}) with $c_1=0$,
we obtain (hereafter we adopt the reduced units
in which $W=k_B=\mu_B=1$)
\begin{eqnarray}
m(T) &=& m(0)+ c_2[\rho'_{\uparrow}-\rho'_{\downarrow}]\:T^2
+\cdot\cdot, 
\label{eq:Y1} \\
n(T) &=& n(0)+ c_2[\rho'_{\uparrow}+\rho'_{\downarrow}]\:T^2
+\cdot\cdot, 
\label{eq:Y2} \\
E(T) &=& E(0)
+ c_2[\rho_{\uparrow}+\rho_{\downarrow}
+\mu\:(\rho'_{\uparrow}+\rho'_{\downarrow})]\:T^2
+\left( \frac{U}{4}\right) m(T)^2+\cdot\cdot, 
\label{eq:Y3}
\end{eqnarray}
where $\rho_{\sigma}=\rho_{\sigma}(\mu)$ and
$\rho'_{\sigma}=d \rho(\mu)/d \epsilon$.
Simple calculations using Eqs. (\ref{eq:Y1})-(\ref{eq:Y3}) lead to 
\begin{eqnarray}
m(T) &=& m(0) 
-\alpha \:T^2, 
\label{eq:G1} \\
C(T) &=& \gamma \:T, 
\label{eq:G2}
\end{eqnarray}
with
\begin{eqnarray}
\alpha &=& c_2 \:(\rho'_{\downarrow}- \rho'_{\uparrow}),
\label{eq:G3} \\
\gamma &=& 2 c_2 (\rho_{\downarrow}+\rho_{\uparrow})
- \alpha U m(0).
\label{eq:G4} 
\end{eqnarray}
The $T^2$-decrease in $m(T)$ arises from the Stoner
excitations. When we take into account spin-wave excitations,
which are neglected in the Hartree-Fock approximation,
magnetization decreases following the $T^{3/2}$ power
at low temperatures.
A rapid decrease in $m(T)$ and a large specific
heat with increasing $\vert q-1 \vert$ shown in
Figs. \ref{fig3} and \ref{fig5}, are attributed to
an enlarged $c_2$ shown in Fig. \ref{fig8}.

Setting $\phi(\epsilon)=d \rho(\epsilon)/d \epsilon$ 
in Eq. (\ref{eq:F1}), we obtain 
\begin{eqnarray}
\chi_{0} &=& \rho + c_2 \rho^{(2)}\:T^2 + c_3 \rho^{(3)}\:T^3
+ \cdot\cdot,
\label{eq:Y4}
\end{eqnarray}
where $\rho^{(\ell)}=\rho^{(\ell)}(\mu)$ ($\ell=2,3$).
The Curie temperature $T_C$ is implicitly given by
$U \chi_{0}(T_C)-1=0$, 
\label{eq:G6}
which yields
\begin{eqnarray}
T_C &=& \left( \frac{U \rho-1}{- c_2 U \rho^{(2)}} \right)^{1/2}.
\label{eq:G7}
\end{eqnarray}
A significant decrease in $T_C$  with increasing  $\vert q-1 \vert$
shown in Fig. \ref{fig4} is again due to an enlarged $c_2$.

For a calculation of the susceptibility at $T > T_C$, 
contributions from
higher terms than $T^2$ are necessary. 
From Eqs. (\ref{eq:E2}) and (\ref{eq:Y4}),
the inverse of the susceptibility is given by
\begin{eqnarray}
\frac{1}{\chi}&=&
\frac{U [c_2 \rho^{(2)}(T^2-T_C^2) 
+ c_3 \rho^{(3)}(T^3-T_C^3) + \cdot \cdot]}
{2[\rho + c_2 \rho^{(2)} T^2 + c_3 \rho^{(3)} T^3+ \cdot \cdot ]}.
\label{eq:G8}
\end{eqnarray}
We note in Eq. (\ref{eq:G8}) that if $c_3=0$, the temperature dependence of
$1/\chi$ becomes symmetric with respect to $q=1.0$
because of a symmetry of $c_2$.
It is not the case because of the significant contribution from $c_3$,
as shown in Fig. \ref{fig6}.

In the present study, we have employed the GFD distribution 
of $f_q(\epsilon)$ given by Eq. (\ref{eq:H10})
obtained within the superstatistics \cite{Wilk00}-\cite{Beck05}.
It is worthwhile to point out that
the resultant GFD distribution depends on a way how the average 
is performed over the fluctuating field in the superstatistics.
Indeed, if taking the average of $f_{BG}(\epsilon_k)$
given by Eq. (\ref{eq:H11})
over the $\chi^2$-distribution, we obtain
\begin{eqnarray}
\tilde{f}_q(\epsilon_k) &=& 
\int_0^{\infty} f_{BG}(\epsilon_k) \:g(\tilde{\beta}) \:d \tilde{\beta}, 
\label{eq:H12} \\
&=& \int_0^{\infty} \frac{1}{1+e^{\tilde{\beta}(\epsilon_k-\mu)}} 
\:g(\tilde{\beta}) \:d \tilde{\beta}.
\label{eq:H13}
\end{eqnarray}
The $\epsilon$ dependence of $\tilde{f}_q(\epsilon_k)$
calculated by numerical methods is shown in Fig. \ref{fig9}.
We note that $\tilde{f}_q(\epsilon_k)$ is rather different from 
$f_q(\epsilon_k)$ given by Eq. (\ref{eq:H10}) except for $q=1.0$.
Fig. \ref{fig9} clearly shows that the average over the
fluctuating field in Eq. (\ref{eq:H13}) leads to result
different from $f_q(\epsilon)$ given by Eq. (\ref{eq:H10}). 
The chain curve in Fig. \ref{fig9} will be explained below.

It is noted that some applications of the nonextensive 
quantum statistics have employed the GFD distribution given by
\cite{Oliveira00,Nunes01} 
\begin{eqnarray}
\bar{f}_q(\epsilon_k) 
&=& \frac{1}{1+(\exp_q[-\beta(\epsilon_k-\mu)])^{-q}}, 
\label{eq:H14}
\end{eqnarray}
in place of $f_q(\epsilon_k)$ in Eq. (\ref{eq:H10}).
A power index $q$ in the denominator of Eq. (\ref{eq:H14}) arises from
the $q$-average of $\langle O \rangle_q = Tr \: \hat{\rho}^q \:\hat{O} $ 
in the MEM (ii),
where $Tr$ denotes the trace, $\hat{\rho}$ the density matrix 
and $\hat{O}$ a given operator.
Such an averaging does not appear in either classical or quantum
superstatistics \cite{Wilk00}-\cite{Beck05}.
The chain curve in Fig. \ref{fig9} shows
the $\epsilon$ dependence of $\bar{f}_q(\epsilon)$ for $q=1.2$, 
which is similar to $f_q(\epsilon)$ shown by the solid curve.

Magnetizations of manganites like 
${\rm La}_{0.60}{\rm Y}_{0.07}{\rm Ca}_{0.33}{\rm MnO}_3$
show a peculiar temperature dependence, which has been shown to
arise from the cut-off properties in the generalized Brillouin
function for nonextensive localized spin systems
\cite{Portesi95}-\cite{Reis06}.
In metallic ferromagnets under consideration, the cut-off properties
appear in the GFD distribution 
[Eq. (\ref{eq:B9})].
Because $m$, $E$ and $C$ are integrated quantities
over the GFD distribution, effects of the cut-off properties
are hardly realized in their temperature dependence.
In both metallic and insulating ferromagnets, the temperature
dependence of $m$ becomes more significant with further
increasing the nonextensivity.
 
When employing Eq. (\ref{eq:G4}) for paramagnetic metals
($\rho_{\downarrow}=\rho_{\uparrow}=\rho$), we obtain 
the $T$-linear coefficient of the specific heat given by
$\gamma = 4 c_2 \rho$, which leads to
the enhancement of the linear coefficient
of the specific heat by the nonextensivity:
\begin{eqnarray}
\frac{\gamma(q)}{\gamma(1)} &=& \frac{c_2(q)}{c_2(1)}.
\label{eq:J1}
\end{eqnarray} 
This ratio is increased with increasing $\vert q-1 \vert$: 
it is 4.28 and 10.46 for $\vert q-1 \vert =0.5$ and 0.9, respectively
[Fig. \ref{fig8}].
Within the superstatistics, this phenomenon may be interpreted
as due to the effect of fluctuating $\beta$ (for $q > 1$).
Similar enhancements in the specific heat
are realized in effects of spin fluctuations at low temperatures
\cite{Berk66,Doniach66} 
and of critical fluctuations near the transition temperatures. 
It has been reported that in some heterogeneous magnetic metals
such as metallic spin glasses and metallic amorphous ferromagnets,
contributions from electronic specific heat are abnormally large
compared to that in normal metals.
The temperature dependence of the specific heat
and susceptibility in nano-magnets has been shown to
considerably depend on the nonextensivity 
\cite{Hasegawa05,Hasegawa07}.
It would be interesting to analyze these materials from the
view point of the nonextensive statistics.

\section{Conclusion}

By using the factorization approach to the GFD distribution \cite{Buy95},
we have discussed magnetic and thermodynamical properties of
nonextensive itinerant-electron ferromagnets
described by the Hubbard model with the Hartree-Fock approximation. 
Our calculation has shown that
with increasing the nonextensivity of $\vert q-1 \vert$, 
Stoner excitations is much increased, 
which induces a significant decrease in the magnetization and
a considerable increase in the specific heat at low temperatures.
The adopted Hartree-Fock approximation has
some disadvantages: it cannot well explain
the $T^{3/2}$-power law of magnetization
at low temperatures,
the large specific-heat anomaly around $T_C$,
and the Curie-Weiss susceptibility.
Nevertheless, the Hartree-Fock approximation has 
an advantage that it provides a reasonable
overall description for magnetic and thermodynamical properties.
For a more accurate  description of  nonextensive quantum systems, it is
necessary to go beyond the factorization approximation to the GFD 
distribution \cite{Hasegawa09}.
Our study may be generalized to various directions:
extensions to ferromagnetic (and antiferromagnetic)
metals and alloys with more complicated structures, and
calculations of various physical quantities such as 
spin waves and conductivity.
We may extend our theory to include effects of
spin fluctuations in nonextensive itinerant-electron ferromagnets 
with the use of the functional-integral method which is useful for
(extensive) bulk ferromagnets \cite{Hasegawa79}\cite{Hubbard79}.

\section*{Acknowledgments}
This work is partly supported by
a Grant-in-Aid for Scientific Research from the Japanese 
Ministry of Education, Culture, Sports, Science and Technology.  

\vspace{0.5cm}
\appendix

\noindent
{\large\bf Appendix: Calculations of $d m/dT$ and $d \mu/dT$}
\renewcommand{\theequation}{A\arabic{equation}}
\setcounter{equation}{0}

The terms of $dm/dT$ and $d \mu/dT$ in Eq. (\ref{eq:D2})
may be derived as follows.
From Eqs. (\ref{eq:C1}) and (\ref{eq:C2}), we obtain 

\begin{eqnarray}
a_{11} \:\left( \frac{d m}{d T}\right) 
+ a_{12} \:\left( \frac{d \mu}{dT} \right) &=& b_1,
\label{eq:X1}\\
a_{21} \:\left( \frac{d m}{d T} \right)
+ a_{22} \:\left( \frac{d \mu}{dT}\right) &=& b_2,
\label{eq:X2}
\end{eqnarray}
with
\begin{eqnarray}
a_{11} &=& 1+ \left( \frac{U}{2} \right) 
\int \:[\rho_{\uparrow }(\epsilon)+\rho_{\downarrow}(\epsilon)] 
\:\frac{\partial f_q(\epsilon)}{\partial \epsilon} 
\: d\epsilon, 
\label{eq:X3} \\
a_{12} &=& 
\int \:[\rho_{\uparrow }(\epsilon)-\rho_{\downarrow}(\epsilon)] 
\:\frac{\partial f_q(\epsilon)}{\partial \epsilon} 
\: d\epsilon, 
\label{eq:X4}\\
a_{21} &=& \left( \frac{U}{2} \right)
\int \:[\rho_{\uparrow }(\epsilon)-\rho_{\downarrow}(\epsilon)] 
\:\frac{\partial f_q(\epsilon)}{\partial \epsilon} 
\: d\epsilon,
\label{eq:X5} \\
a_{22} &=& 
\int \:[\rho_{\uparrow }(\epsilon)+\rho_{\downarrow}(\epsilon)] 
\:\frac{\partial f_q(\epsilon)}{\partial \epsilon} 
\: d\epsilon, 
\label{eq:X6}\\
b_{1} &=& - \frac{1}{T} \int \:(\epsilon-\mu)
\:[\rho_{\uparrow }(\epsilon)-\rho_{\downarrow}(\epsilon)] 
\:\frac{\partial f_q(\epsilon)}{\partial \epsilon} 
\: d\epsilon, 
\label{eq:X7} \\
b_{2} &=& - \frac{1}{T} \int \:(\epsilon-\mu)
\:[\rho_{\uparrow }(\epsilon)+\rho_{\downarrow}(\epsilon)] 
\:\frac{\partial f_q(\epsilon)}{\partial \epsilon} 
\: d\epsilon, 
\end{eqnarray}
By solving Eqs (\ref{eq:X1}) and (\ref{eq:X2}), we obtain
\begin{eqnarray}
\frac{d m}{d T} &=& \frac{(a_{22}b_1-a_{12} b_2)}{\Delta}, 
\label{eq:X9} \\
\frac{d \mu}{d T} &=& \frac{(-a_{21}b_1+a_{11} b_2)}{\Delta},
\label{eq:X10}
\end{eqnarray}
where $\Delta=a_{11}a_{22}-a_{12}a_{21}$.

Coefficients given by Eq. (\ref{eq:X3})-(\ref{eq:X6}) are used also for
solving Eqs. (\ref{eq:C1})-(\ref{eq:C2}) by the Newton-Raphson method.
The $\ell$th iterative solution of
$m_{\ell}$ and $\mu_{\ell}$ of Eqs. (\ref{eq:C1}) and (\ref{eq:C2}) are
given by
\begin{eqnarray}
m_{\ell} &=& m_{\ell-1}
+\frac{(a_{22}c_1-a_{12} c_2)}{\Delta}, 
\label{eq:X11}\\
\mu_{\ell} &=& \mu_{\ell-1}
+\frac{(-a_{21}c_1+a_{11} c_2)}{\Delta},
\label{eq:X12}
\end{eqnarray}
with
\begin{eqnarray}
c_1 &=& \int [\rho_{\uparrow}(\epsilon)-\rho_{\downarrow}(\epsilon)]
f_q(\epsilon) \;d \epsilon - m_{\ell-1}, 
\label{eq:X13} \\
c_2 &=& \int [\rho_{\uparrow}(\epsilon)+\rho_{\downarrow}(\epsilon)]
f_q(\epsilon) \;d \epsilon -n,
\label{eq:X14}
\end{eqnarray}
where the first terms in Eqs. (\ref{eq:X13}) and (\ref{eq:X14}) are
expressed in terms of the $(\ell-1)$th solutions 
of $m_{\ell-1}$ and $\mu_{\ell-1}$.

\begin{figure}
\begin{center}
\end{center}
\caption{
(Color online)
The energy dependence of (a) the generalized Fermi-Dirac distribution
$f_q(\epsilon)$ 
and (b) its derivative of $- \partial f_q(\epsilon)/\partial \epsilon$
for various $q$ with $\beta=1.0$.
}
\label{fig1}
\end{figure}

\begin{figure}
\begin{center}
\end{center}
\caption{
The $U$ dependence of the magnetic moment $m$ 
for $k_B T/W=0.0$, $\mu_B B/W=0.0$
and $n=1.4$ electrons, the arrow denoting the $U/W$ value (=1.75)
adopted in model calculations.
}
\label{fig2}
\end{figure}

\begin{figure}
\begin{center}
\end{center}
\caption{
(Color online)
The temperature dependence of the magnetic moment $m$
for (a) $q \leq 1.0$ and (b) $q \geq 1.0$ with $\mu_B B/W=0.0$.
}
\label{fig3}
\end{figure}

\begin{figure}
\begin{center}
\end{center}
\caption{
The Curie temperature $T_C$ as a function of $q$.
}
\label{fig4}
\end{figure}

\begin{figure}
\begin{center}
\end{center}
\caption{
(Color online)
The temperature dependence of the specific heat $C$
for (a) $q \leq 1.0$ and (b) $q \geq 1.0$ with $\mu_B B/W=0.0$.
}
\label{fig5}
\end{figure}

\begin{figure}
\begin{center}
\end{center}
\caption{
(Color online)
The temperature dependence of the inversed susceptibility $1/\chi$
for (a) $q \leq 1.0$ and (b) $q \geq 1.0$,
insets showing the enlarged plots for $0.0 \leq T \leq 0.4$.
}
\label{fig6}
\end{figure}

\begin{figure}
\begin{center}
\end{center}
\caption{
(Color online)
The $U$ dependence of the Curie temperature $T_C$ 
for $q=1.0$ (the dashed curve), $q=0.2$ (the chain curve)
 and $q=1.8$ (the solid curve).
}
\label{fig7}
\end{figure}

\begin{figure}
\begin{center}
\end{center}
\caption{
(Color online)
The $q$ dependence of the coefficients $c_n$
in the generalized Sommerfeld expansion [Eq. (\ref{eq:F1})]:
$c_1$ (the dashed curve), $c_2$ (the solid curve),
$c_3$ (the chain curve) and $c_4$ (the double-chain curve).
}
\label{fig8}
\end{figure}

\begin{figure}
\begin{center}
\end{center}
\caption{
(Color online)
The energy dependence of the GFD distributions of 
$f_q(\epsilon)$ [Eq. (\ref{eq:H10})](the solid curve),
$\tilde{f}_q(\epsilon)$ [Eq. (\ref{eq:H13})] (the dashed curve)
and $\bar{f}_q(\epsilon)$ [Eq. (\ref{eq:H14})] (the chain curve)
for $q=1.2$: the three GFD distributions agree for $q=1.0$
(the dotted curve) (see text).
}
\label{fig9}
\end{figure}

\end{document}